\documentstyle[aps,multicol,epsfig]{revtex}

\begin{document}
\draft
\title{The Density Matrix Renormalization Group technique with periodic
boundary conditions}
\author{A.~Gendiar and A.~\v{S}urda}
\address{Institute of Physics, Slovak Academy of Sciences,\\ 
D\'{u}bravsk\'{a} cesta 9, SK-842 28 Bratislava, Slovak Republic}
\date \today
\maketitle

\begin{abstract}
The Density Matrix Renormalization Group (DMRG) method with periodic
boundary conditions is introduced for two dimensional classical spin
models. It is shown that this method is more suitable for derivation of
the properties of infinite 2D systems than the DMRG with open boundary
conditions despite the latter describes much better strips of finite width.
For calculation at criticality, phenomenological renormalization at finite
strips is used together with a criterion for optimum strip width
for a given order of approximation. For this width the critical temperature
of 2D Ising model is estimated with seven-digit accuracy for not too large
order of approximation. Similar precision is reached for critical
indices. These results exceed the accuracy of similar calculations for DMRG 
with open boundary conditions by several orders of magnitude.
\end{abstract}

\pacs{05.70.Fh, 64.70.Rh, 02.60.Dc, 02.70.-c}

\widetext

\renewcommand{\theequation}{\arabic{equation}}
\setcounter{equation}{0}
\section{Introduction}

In 1992 the density matrix renormalization group (DMRG) technique in real 
space was invented by S. R. White \cite{Whi} and has been mostly applied to the 
diagonalization of one-dimensional (1D) quantum chain spin Hamiltonians. 
Three years later the DMRG method was redesigned by T. Nishino \cite{Nis} 
and applied to classical spin 2D models. The DMRG method for the classical 
models is based on renormalization of the transfer matrix. It is a variational 
method maximizing the partition function using a limited number of degrees of 
freedom and its variational state is written as a product of local matrices 
\cite{Var,Sur}.

DMRG has been used for many various quantum models. It provides 
results with remarkable accuracy for larger systems than it is possible
to study using standard diagonalization methods. The 2D classical systems treated by 
the DMRG method exceeds the classical Monte Carlo approach in accuracy, 
speed, and size of the systems \cite{Kar}. 
A further DMRG improvement of the classical systems is based 
on Baxter's corner transfer matrix \cite{CTM}, the CTMRG 
\cite{CTMRG}, and its generalization to any dimension \cite{3DN}.

Applications of the DMRG technique for calculation of the thermodynamic
 properties in the
2D classical systems has been done by \cite{Nis,Drw,Ni2,Hon}. Treating of  
non-symmetric transfer matrices or non-hermitian quantum Hamiltonians has
also been studied by the DMRG technique \cite{Hie,Non,Car,And}.

It was shown that DRMG method yields very accurate estimations of ground state
energy of finite quantum chains and free energy of classical strips  of finite width 
with open boundary conditions. We have developed the DMRG method with periodic boundary
conditions for strips of classical spins and shown that, similarly as for quantum
chains, it gives these quantities with much less degree of accuracy.  Nevertheless, 
the DMRG method is mostly used for prediction of physical quantities and
critical properties of infinite  systems in connection with finite size
scaling or extrapolation of the results from finite-size systems to infinite
ones. The objective of this paper is to study the DMRG and exact methods with two
different boundary conditions for finite strips of various widths and compare 
their results with known exact results for infinite 2D system. It is shown
that while for the exact diagonalization of finite-strip transfer matrices
scaling properties of the system improve, for DMRG approach there exists an
optimum width for each degree of approximation. The developed approach is
tested on 2D Ising model. 

Paper is organized as follows: in Sec.~II  we  mention briefly  the DMRG
for the open boundary conditions; Sec.~III contains the modification of the
DMRG method for the periodic boundary conditions; in Sec.~IV we present
the results obtained by the DMRG method with periodic and open boundary
conditions, the exact diagonalization method and show how to determine the 
optimum strip width for finite-size scaling, and in Sec.~V the results will
be summarized.

\section{DMRG with open boundary conditions}

The transfer matrix approach is a powerful method for exact numerical 
calculation of
thermodynamical properties of lattice spin models defined on finite-width
strips. If the width of the strip is too large and the capacity of the
computer is exceeded, the DMRG method is found to be useful for an
effective  reduction
of the transfer matrix size. It can be used for calculation of global
quantities such as free energy as well as of a spatial dependence  across the
strip  of local
quantities, e.g. spin correlation functions. 

The properties of an infinite strip of finite width $L$ are given by the
solution of `left' eigenvectors and corresponding eigenvalues of the transfer matrix equation    
\begin{equation}
\sum\limits_{\{\sigma\}}\Psi_l\left(\{\sigma\}\right)T\left(\{\sigma\}|\{\sigma^{\prime}\}\right)=
\lambda\Psi_l\left(\{\sigma^{\prime}\}\right),
\label{eigeq}
\end{equation}
where  $\{\sigma\}$ is a set of $L$ spins 
$\{\sigma_1,\sigma_2,\dots,\sigma_L\}$
defined on a row  and $\{\sigma^{\prime}\}$ is a set of $L$ spins 
on the adjacent row.
The transfer matrix is a product of  Boltzmann weights given by the lattice
Hamiltonian. For non-symmetric transfer matrices besides the left
eigenvectors  
$\Psi^i_l$, the right eigenvectors $\Psi^i_r$ should be calculated, as well.

Reducing the size of the transfer matrix the standard DMRG technique 
proceeds in two regimes:
\newline
(1) In the process of iterations, the infinite system method (ISM) pushes 
both ends of the transfer matrix further so that each step
of the ISM enlarges the lattice size by two sites. The transfer matrix 
(superblock) is constructed
from three blocks: left $T_l$ and right $T_r$ transfer matrices (blocks) 
and a Boltzmann weight $W_B$,
in particular
\begin{equation}
T_{[2j+2]}=T^{(j)}_lW_BT^{(j)}_r,
\end{equation}
where the index on the left hand side denotes the number of sites in one row 
of the whole superblock $T$ 
at the $j$th step of iteration.
The Boltzmann weight usually is a function  of several spins interacting among 
each other, e.g. for  the
Ising model with nearest-neighbor interactions the Boltzmann weight has the form
\begin{equation}
W_B(\sigma_1\sigma_2|\sigma_1^{\prime}\sigma_2^{\prime})=\exp\left\{-\frac{J}{k_BT}(\sigma_1\sigma_2
+\sigma_1^{\prime}\sigma_2^{\prime}+\sigma_1\sigma_1^{\prime}+\sigma_2\sigma_2^{\prime})\right\}.
\end{equation}

In the first step of the ISM (for details, see \cite {Whi,Nis}) 
 $T_l^{(1)}=T_r^{(1)}=W_B$ is put. 
The whole procedure has $L/2-1$ steps
\begin{equation}
T_l^{(1)}W_BT_r^{(1)}\rightarrow T_l^{(2)}W_BT_r^{(2)}\rightarrow\cdots\rightarrow 
T_l^{(L/2-1)}W_BT_r^{(L/2-1)}. 
\label{ism}
\end{equation}
and stops when  the desired strip width of $L$ sites is reached.

The first steps of the iteration scheme (\ref{ism}) are exact
but if the superblock matrix $T$ becomes too large, a reduction procedure,
to keep the size of superblock constant,
should    be introduced.

 The first step  of (4) introduces open 
conditions at the   strip boundaries. If the temperature of the system is
lower than the critical one and the strip width is wide enough, the symmetry
of the system is  spontaneously broken (order parameter becomes non-zero),
and after reaching the fixed point of the iteration procedure, the system
does not depend on the boundary conditions any more. The calculations with  
periodic boundary 
conditions
described in the next Section give in this regime the same result as with the free ones.
\newline
(2) the finite system method (FSM) improves numerical accuracy of ISM result  by left and right moves
({\it sweeps}) according to the following prescription:
\begin{equation}
T_l^{(L/2-1)}W_BT_r^{(L/2-1)}\rightarrow T_l^{(L/2)}W_BT_r^{(L/2-2)}\rightarrow\cdots\rightarrow 
T_l^{(L-3)}W_BT_r^{(1)},
\label{fsm1}
\end{equation}
\begin{equation}
T_l^{(L-3)}W_BT_r^{(1)}\rightarrow T_l^{(L-2)}W_BT_r^{(2)}\rightarrow\cdots\rightarrow 
T_l^{(L/2-1)}W_BT_r^{(L/2-1)}. 
\label{fsm2}
\end{equation}

In the right sweep (5) the left blocks $T_l$ are calculated in the previous
step of the sweep and the right blocks $T_r$ are taken from the previous
left sweep (in the first right sweep from ISM); similarly for the
left sweep.
 
The values of local thermodynamical quantities given by particular superblocks in
the final sweep (after the steady state is reached) are spatially dependent.
The values given by the superblock in the middle of the strip 
 $T_l^{(L/2-1)}W_BT_r^{(L/2-1)}$ 
are the closest to the bulk ones. In this sense, the best transfer matrix eigenvalues 
as well as eigenvectors are those of the above-mentioned central superblock. 
The two largest eigenvalues are used for further finite-size 
scaling or extrapolation treatment.

\section{DMRG with periodic boundary conditions}

The translational invariance of the infinite lattice is preserved in finite
strips with periodic boundary conditions when strip boundaries are connected
with bulk intersite interactions. In this case the strip forms 
an infinitely long cylinder. If the radius of the cylinder is small enough,
the model can be easily solved by exact numerical diagonalization methods. 

In DMRG language, imposing periodic boundary conditions means that 
we have to introduce properly the connection of both ends of the superblock 
transfer matrix $T$. 
Thus, in distinction to open-boundary case the superblock is constructed from
two Boltzmann weights connecting two blocks at both ends (see Fig.~1,
the rightmost diagram).
\begin{eqnarray}
\nonumber
T_{[2j+4]}(\sigma_1\xi_l\sigma_{j+4}\sigma_{j+3}\xi_r\sigma_2|\sigma_1^{\prime}\xi_l^{\prime}
\sigma_{j+4}^{\prime}\sigma_{j+3}^{\prime}\xi_r^{\prime}\sigma_2^{\prime})=T_l^{(j)}(\sigma_1
\xi_l\sigma_{j+4}|\sigma_1^{\prime}\xi_l^{\prime}\sigma_{j+4}^{\prime})\\
\times
W_B(\sigma_{j+4}\sigma_{j+3}|\sigma_{j+4}^{\prime}\sigma_{j+3}^{\prime})T_r^{(j)}(\sigma_{j+3}
\xi_r\sigma_2|\sigma_{j+3}^{\prime}\xi_r^{\prime}\sigma_2^{\prime})W_B(\sigma_2\sigma_1|
\sigma_2^{\prime}\sigma_1^{\prime}),
\end{eqnarray}
where the block spin variable $\xi=\{1,2,\dots,m\}$, and the primed variables
are denoted  by filled circles and ovals in Fig.~1.

In the first few steps the lattice is enlarged to the desired size; no 
degrees-of-freedom reduction is performed and the superblock transfer
matrix remains equivalent to the exact one. As depicted in Fig.~1, 
the ISM starts with $T^{(6)}=T_l^{(1)}W_BT_r^{(1)}W_B$ defined on 
twelve sites  where $T_l^{(1)}=T_r^{(1)}=W_BW_B$, and 
one Boltzmann weight, i.e. four new sites are added in each of further steps.   

\begin{figure}[tb]
\centering
\centerline{\scalebox{0.55}{\includegraphics{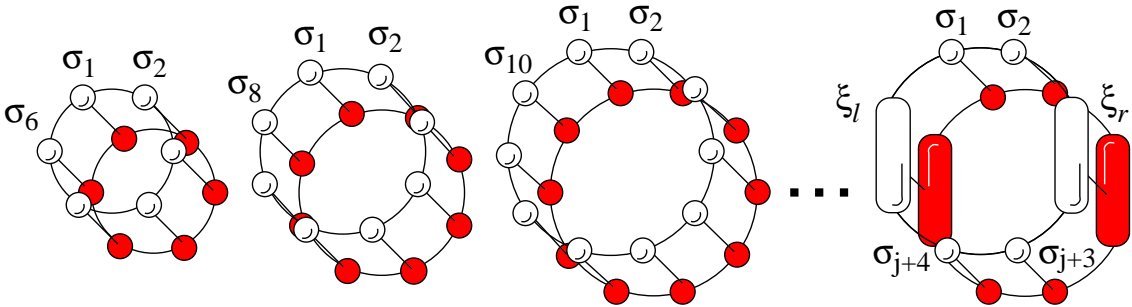}}}
\caption{\it The first $j$ steps of the ISM for the strip with the periodic boundary
conditions.}
\label{3DviewofISM}
\end{figure}

If $2^j>m$, the number of degrees of freedom should be
reduced at each $j$th step to keep the order of the superblock matrix constant and 
equal to $2^4\cdot m^2$.

\begin{figure}[tb]
\centering
\centerline{\scalebox{0.55}{\includegraphics{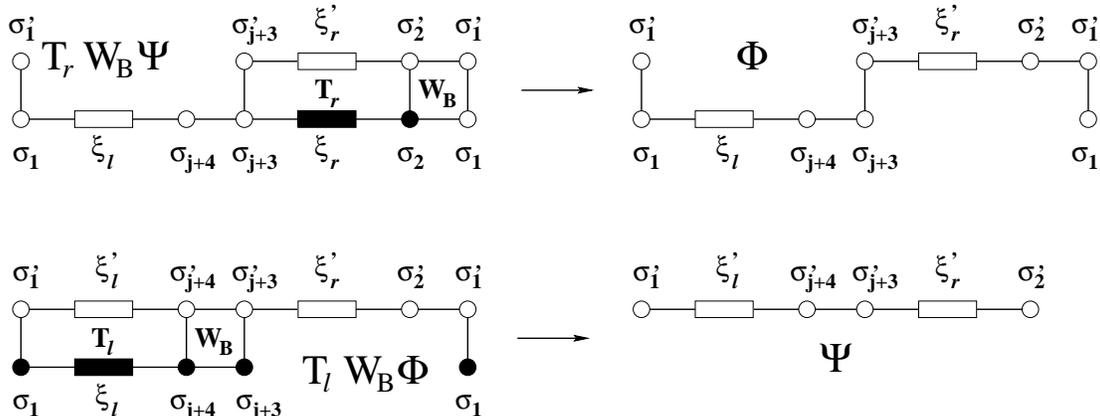}}}
\caption{\it Graphical representation of Eq.~(\ref{eqprb}). 
The variables represented by filled circles and rectangles are summed over.
The spins $\sigma_1$ and $\sigma_1^{\prime}$ at both ends of the superblock
 must be identified due to the periodic boundary conditions.} 
\label{graphprb}
\end{figure}

Summation in the equation for eigenvectors (1) of the transfer matrix (8) can be
performed in two steps
\begin{eqnarray}
\nonumber
\Phi(\sigma_1\xi_l\sigma_{j+4}\sigma_{j+3}\sigma_{j+3}^{\prime}\xi_r^{\prime}\sigma_2^{\prime}
\sigma_1^{\prime}) &
= &
\sum\limits_{\xi_r\sigma_2}T_r(\sigma_{j+3}\xi_r\sigma_2|\sigma_{j+3}^{\prime}\xi_r^{\prime}
\sigma_2^{\prime})W_B(\sigma_2\sigma_1|\sigma_2^{\prime}\sigma_1^{\prime})\\
\nonumber
& &
\times\Psi(\sigma_1\xi_l\sigma_{j+4}\sigma_{j+3}\xi_r\sigma_2)\\
\Psi(\sigma_1^{\prime}\xi_l^{\prime}\sigma_{j+4}^{\prime}\sigma_{j+3}^{\prime}
\xi_r^{\prime}\sigma_2^{\prime}) &
= &
\sum\limits_{{\sigma_1\xi_l}\atop{\sigma_{j+4}\sigma_{j+3}}}T_l(\sigma_1\xi_l\sigma_{j+4}|
\sigma_1^{\prime}\xi_l^{\prime}\sigma_{j+4}^{\prime})W_B(\sigma_{j+4}\sigma_{j+3}|
\sigma_{j+4}^{\prime}\sigma_{j+3}^{\prime})\\
\nonumber
& &
\times\Phi(\sigma_1\xi_l\sigma_{j+4}\sigma_{j+3}\sigma_{j+3}^{\prime}\xi_r^{\prime}
\sigma_2^{\prime}\sigma_1^{\prime})
\label{eqprb}
\end{eqnarray}
which is depicted  graphically  in Fig.~\ref{graphprb}.

This procedure uses the left and right transfer matrix blocks to calculate
properly the left and right 
eigenvectors $\Psi_l$ and $\Psi_r$, respectively, of the whole superblock for the 
periodic boundary conditions.
Once we have the $\Psi_l$ and $\Psi_r$, the left and right 
density matrices  can be constructed. 
\begin{equation}
\rho_l(\xi_l\sigma_{j+4}|\xi_l^{\prime}\sigma_{j+4}^{\prime})=\sum\limits_{\sigma_1\sigma_{j+3}
\xi_r\sigma_2}\Psi_l(\sigma_1\xi_l\sigma_{j+4}\sigma_{j+3}\xi_r\sigma_2)
\Psi_r(\sigma_1\xi_l^{\prime}\sigma_{j+4}^{\prime}\sigma_{j+3}\xi_r\sigma_2)
\end{equation}
\begin{equation}
\rho_r(\sigma_{j+3}\xi_r|\sigma_{j+3}^{\prime}\xi_r^{\prime})=\sum\limits_{\sigma_1\xi_l
\sigma_{j+4}\sigma_2}\Psi_l(\sigma_1\xi_l\sigma_{j+4}\sigma_{j+3}\xi_r\sigma_2)
\Psi_r(\sigma_1\xi_l\sigma_{j+4}\sigma_{j+3}^{\prime}\xi_r^{\prime}\sigma_2),
\end{equation}
and by its complete diagonalization 
\begin{equation}
O_l(\xi_l^{new}|\xi_l\sigma_{j+4})\rho_l(\xi_l\sigma_{j+4}|\xi_l^{\prime}\sigma_{j+4}^{\prime})
Q_l(\xi_l^{\prime}\sigma_{j+4}^{\prime}|\xi_l^{\prime\ new})=\omega_i\delta_{ij}
\end{equation}
sets of left and right eigenvectors stored in $O_l$ and $Q_l$ matrices,
respectively, is obtained (analogously, for $O_r$ and $Q_r$). 
The indices $i,j$ ($i,j\!=\!1,2,\dots,2m$) run over all states of 
$m$-state multi-spin variable $\xi$ and two-state variable $\sigma$.   
For the last steps of ISM and all FSM steps
half of the eigenvectors (corresponding to their lowest eigenvalues) is
discarded from 
the matrices $O$ and $Q$, and the  information of the system carried by the density matrix
is reduced.
However, remaining eigenvectors (if $m$ is large enough) usually describe the system accurately 
because the truncation error $\varepsilon$ defined as
\begin{equation}
\varepsilon=\sum\limits_{\{{\rm discarded}\}}\omega
\label{trunc-err}
\end{equation}
is very small ($0\leq\varepsilon\ll 1$). 
$\sum_{\{all\}}\omega\!=\!1$, as the eigenvectors  are assumed to be
normalized. 
The matrices $O$ and $Q$ enter the linear transformation as projectors mapping
two blocks $T_lW_B$ onto one block $T_l$ through the following procedure
\begin{eqnarray}
\nonumber
T_l^{(j+1)}(\sigma_1\xi_l^{new}\sigma_{j+5}|\sigma_1^{\prime}\xi_l^{\prime\ new}\sigma_{j+5}
^{\prime})=\sum\limits_{{\xi_l^{\prime}\sigma_{j+4}^{\prime}}\atop{\xi_l\sigma_{j+4}}} & O_l
(\xi_l^{new}|\xi_l\sigma_{j+4})T_l^{(j)}(\sigma_1\xi_l\sigma_{j+4}|
\sigma_1^{\prime}\xi_l^{\prime}\sigma_{j+4}^{\prime})\\
\times & W_B(\sigma_{j+4}\sigma_{j+5}|\sigma_{j+4}^{\prime}\sigma_{j+5}^{\prime})
Q_l(\xi_l^{\prime}\sigma_{j+4}^{\prime}|\xi_l^{\prime\ new}).
\end{eqnarray}
Application to the right block $T_r$ is straightforward. As it is seen, we
calculate 
 the blocks $T_l$ and $T_r$ separately  not using the
standard mirror-reflection
of $T_l$ to $T_r$.  This procedure is necessary when dealing with anisotropic 
and/or inhomogeneous systems.

The calculated new blocks $T_l^{(j+1)}$ and $T_r^{(j+1)}$ are used in the
next step of the ISM for construction of the new superblock
\begin{equation}
T_{[2j+6]}=T_l^{(j+1)}W_BT_r^{(j+1)}W_B.
\end{equation}

Within the FSM, e.g., for a sweep to the right  only the left  blocks
are calculated and  $T_r^{(L/2-k)}$ is taken from the previous left sweep
\begin{equation} 
T_{[L]}=T_l^{(L/2-3+k)}W_BT_r^{(L/2-1-k)}W_B.
\end{equation}
The variable $k$ (indexing the steps within a sweep) runs over the values
$(-k_0, -k_0+1, \dots, k_0-1, k_0)$, where $2^{k_0}\le 2^{L/2-2}-m$.  
In the process of sweeping one of the Boltzmann weight is fixed (the upper
one in Fig.~1) and the second one changes its position within the interval
of
$2k_0$ lattice sites. The local physical quantities are calculated at the
lattice sites of the fixed Boltzmann weight and due to the rotational
invariance of the problem are valid for all the rows of the periodic lattice.

\section{Results}

It is well known that the DMRG  describes
better a strip with open boundary conditions than that with the  periodic
boundary conditions \cite{Whi} because the precision of the largest 
eigenvalue of the superblock matrix is
increasing proportionally  to $m$ for open boundary conditions while for
periodic boundary conditions only as $\sqrt{m}$.

\begin{figure}[tb]
\caption {\it Free energy per site $f_{\rm ISM}$ for the Ising model calculated 
with the standard DMRG 
method only with the ISM is compared with the free energy per site calculated 
by the modified DMRG algorithm as well as by the exact diagonalization method (EDM).
$N$ is the order of either the superblock of DMRG or the exact transfer matrix in EDM.}
\vskip 0.2truecm
\centering
\centerline{\scalebox{1.0}{\includegraphics{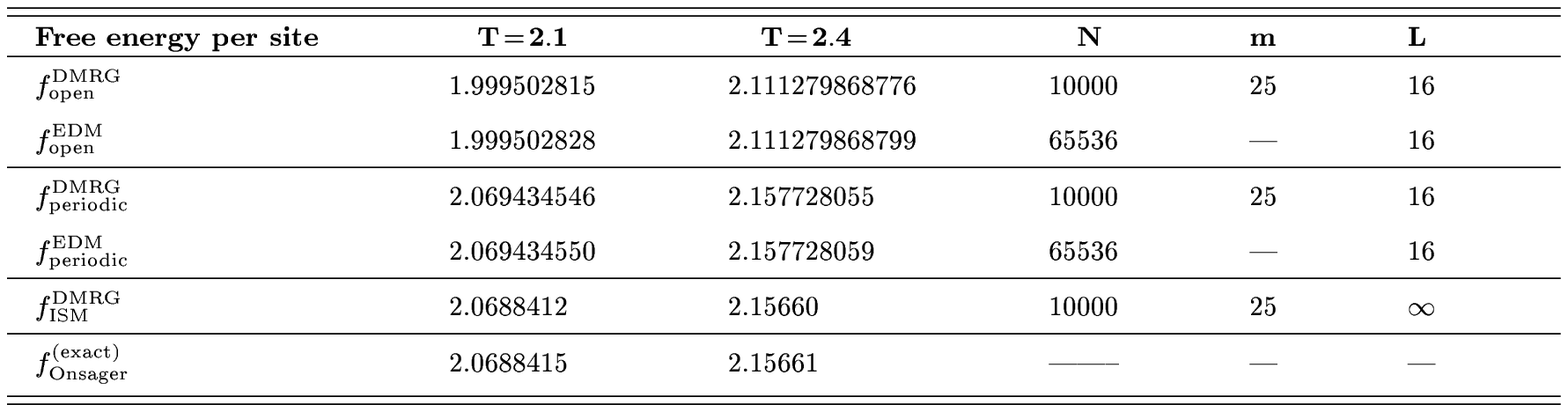}}}
\label{FE}
\end{figure}

However, if we are not interested in the largest eigenvalue of a 
finite-strip transfer matrix but in the estimation
of the free energy of the whole 2D lattice (per spin), it is more effective
to use a strip with periodic boundary conditions than that with open boundaries, as
demonstrated  in  Table~\ref{FE}. The results with $m=25$ practically
exactly reproduce the  exact values for $L=16$. 
The estimation of the free energy for 2D models performed by DMRG can be improved
by increasing the width of the strip. For a given $m$ the best results are obtained
for $L\rightarrow\infty$, but in this case,
for $T=2.1$ (i.e. below the critical temperature), the symmetry of the system
is spontaneously broken. Exact free energy per site
$f^{\rm (exact)}_{\rm Onsager}$ was taken from \cite{Lan}.

The critical temperature and the properties of the infinite 2D system near the critical
temperature should be derived from finite-size scaling ideas (as the finite-width 
strip is at criticality for $T=0$ only). 

For calculation of the critical temperature, the phenomenological renormalization 
approach of Nightingale \cite{Nig} have been used. Here, the scaling
properties of the correlation length, found as  logarithm of the ratio of
two largest eigenvalues of the exact or superblock matrix, are exploited.
The product of the inverse correlation length $K_{N}$ and the strip width $L$ should
not depend on the $L$ at critical temperature $T_{\rm C}^*(L)$
\begin{equation}
\frac{LK_{L}}{(L+2)K_{L+2}}=1.
\end{equation}

The accuracy of the approximate critical temperature improves with size of
the strip in the case of exact diagonalization. For DMRG calculations this
statement is no longer valid, as for very large $L$ the symmetry of the system
spontaneously breaks, and the phenomenological renormalization is not
applicable any more.  Thus, for given order of  approximation $m$, there
exists an optimum value of the strip width $L^{\rm opt}$.  This can be estimated from
the following considerations: For exact diagonalization  or DMRG calculations
 with $m$ close to $2^{L/2-2}$, the difference of the approximate critical temperature 
from the exact critical temperature
$T_{\rm C}^{\rm (exact)}=2\ln^{-1}(1+\sqrt{2})$ \cite{Bax} scales
with the width of the strip as follows \cite{Bar}:
\begin{equation}
\frac{T_{\rm C}^*(L) - T_{\rm C}^{(\rm exact)}}{T_{\rm C}^{(\rm exact)}} \sim
L^{-1/\nu},
\end{equation}
i.e. the ratio
\begin{equation}
\label{scal}
R\equiv \frac{\frac{{\rm d}}{{\rm d}L} T_{\rm C}^*(L)}{\frac{{\rm d}^2}
{{\rm d}L^2}T_{\rm C}^*(L)} =\frac{\nu}{\nu +1}L\sim L.  
\end{equation}

The optimum width $L^{\rm opt}$ should be less than $L_{\rm C}$
for which the ratio of the derivatives $R(L_{\rm C})$ (\ref{scal}) is substantially 
deviated from the originally linear behavior.
In our calculations we have considered the DMRG results to be incorrect for
$R=0$ or $\infty$. In the case of $R=0$, the precise value of $L^{\rm opt}$ is
not too important as the first derivative or change of $T_{\rm C}^*(L)$ is very
small. Near $R=\infty$, a sharp drop of the second derivative of $T_{\rm
C}^*(L)$ to zero is required; indeed, the change of the distance from the line
${\nu\over \nu +1}L$ by more than one order of magnitude takes place within one step of
strip-width enlargement. 

\begin{figure}[tb]
\centering
\centerline{\scalebox{0.6}{\includegraphics{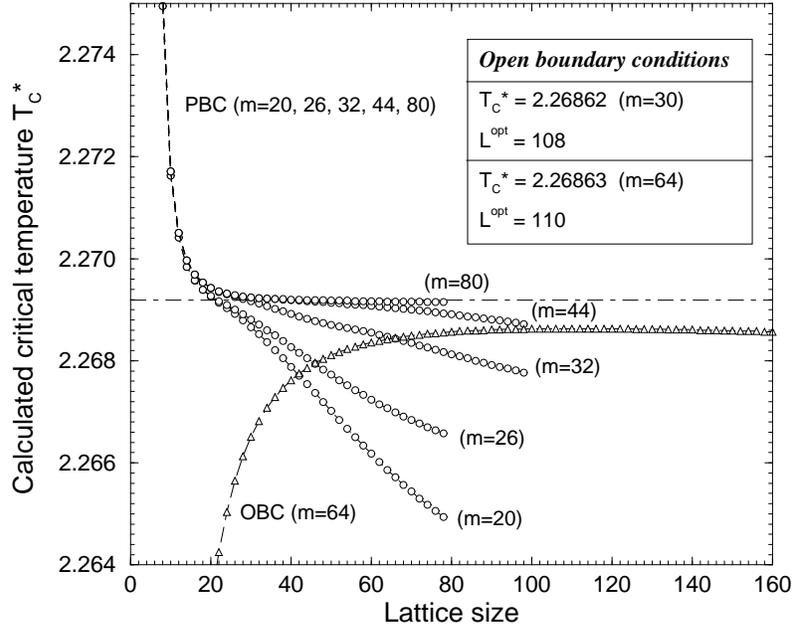}}}
\caption{\it Critical temperatures $T_{\rm C}^*$ for Ising model as functions of lattice size 
$L$ for various sizes of multi-spin variables $m$ from DMRG and finite-size scaling. The
results for open boundary conditions (OBC) are plotted as triangles while the results 
for the periodic boundary conditions (PBC) are plotted as circles. The exact critical
temperature is at the dot-dashed line. The OBC plot for $m=30$ is indistinguishable from
the curve for $m=64$ in this figure.}
\label{PBC_OBC}
\end{figure}

In Fig.~\ref{PBC_OBC} plots of strip-width-dependent critical temperatures $T_{\rm C}^*(L)$
for two different boundary conditions and various block sizes $m$ are given. The
estimations of the exact critical temperature for periodic and open boundary
conditions  were found as the values of $T_{\rm C}^*(L-2)$ 
if the first or second derivative of $T_{\rm C}^*(L)$
changed their signs with respect to the value in the previous step.
The curves for PBC cross the exact
value of $T_{\rm C}^{\rm (exact)}$. The curve maxima for OBC are quite far from it, and by 
increasing $L$, $T_{\rm C}^*$ approaches the exact value very slowly.

\begin{figure}[tb]
\centering
\centerline{\scalebox{0.6}{\includegraphics{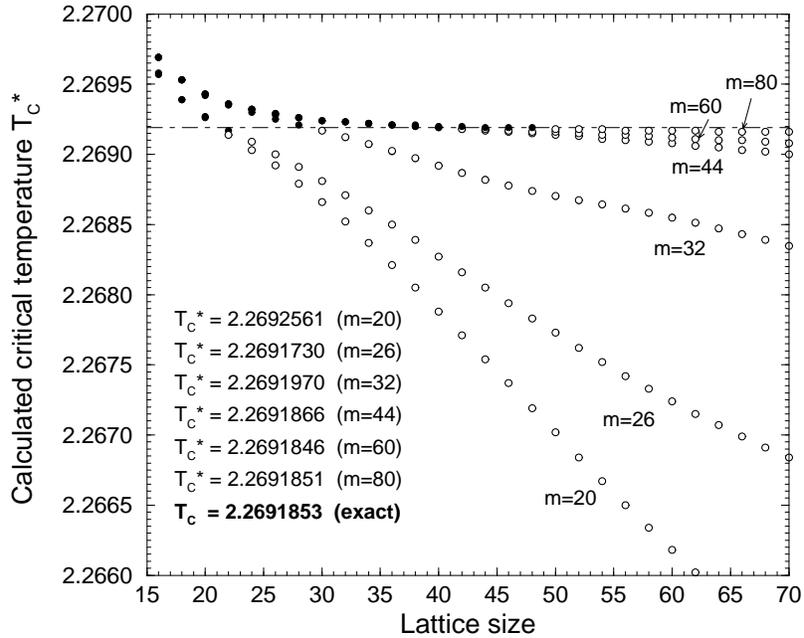}}}
\caption{\it Critical temperatures $T_{\rm C}^*$ in Ising model vs. lattice size $L$ and $m$ for
the periodic boundary conditions (PBC) only. Filled circles represents data which
are accepted whereas the open circles are taken as incorrect due to violation of the condition 
(\ref{scal}). Zoomed Fig.~\ref{PBC_OBC}. The critical temperature estimations in the inset
are given by the rightmost filled circles for respective $m$.}
\label{PBC}
\end{figure}

\begin{figure}[tb]
\caption {\it Changes of $T_{\rm C}^*(L)$ per one step of strip-width enlargement as 
well as the deviations of our results from
the exact critical temperature $T_{\rm C}^{\rm (exact)}$ for increasing parameter $m$.}
\vskip 0.2truecm
\centering
\centerline{\scalebox{1.0}{\includegraphics{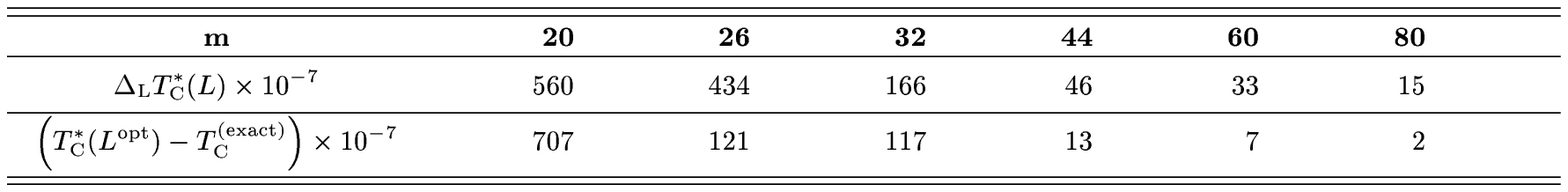}}}
\label{Tc}
\end{figure}

The accuracy of the results for periodic boundary conditions (Fig.~\ref{PBC})
is very high already at small values of $m$ and exceeds by an order the critical
temperature estimation for maximum computer-accessible $m$ when using open boundary
conditions. The critical temperature for not extremely large $m=80$ is
given to seven digits. As the width of the strip can be increased only in
discrete steps and the criterion of the optimum strip-width is somewhat
vague,  the accuracy of the  critical temperature determination should be
taken as large as a single step change of $T_{\rm C}^*(L)$. These accuracy
estimations together with  deviations of our results from the exact critical
temperature are given in Table \ref{Tc}. It should be noted that only ISM was
performed in calculations of $T_{\rm C}^*(L)$ in Figs.~\ref{PBC_OBC} and \ref{PBC}. The
calculations with the FSM has been also done
near the $L^{\rm opt}$ but only slight improvements of critical temperature were obtained. 
In calculation of the thermal critical exponent \cite{Nig} a similar accuracy was
reached; e.g., for $m=44$ within FSM the critical exponent $\nu= 1.0000016$ is very
close to the exact value $\nu=1$ \cite{Bax}.

\section{Conclusion}

The DMRG method for classical spin lattice strips with periodic boundaries
was developed and applied to 2D Ising model. It was shown that this approach
lead to more accurate results for 2D infinite lattice than DMRG with open
boundary conditions. It was demonstrated that applying  finite size scaling
to strips treated by DMRG, an optimal width of the strip depending on the
order of approximation existed, and a prescription how to find $T_{\rm C}^*(L^{\rm opt})$ 
was given. For the Ising model it was shown by computations that for  these  $L^{\rm opt}(m)$  
the value of the critical temperature, was for a given $m$ closest to the exact one. As our
approach does not involve any information about the exact critical temperature 
$T_{\rm C}^{\rm (exact)}$ or the universality class of the model, we believe that it is 
applicable to many different classes of spin lattice models.
This belief is supported by analogous calculation
for anisotropic triangular nearest-neighbor Ising model (ATNNI) with two different
antiferromagnetic interactions $J_1$ and $J_2$ (model discussed in \cite{And}).
For this model the transfer matrix is non-symmetric and the phase diagram is quite different
from that of the standard Ising model. For the periodic boundary conditions, the plot of
critical temperatures is not monotonously decreasing as in the case of the Ising model
(Fig.~\ref{PBC}) but for large $L$ it turns up. Nevertheless, the accuracy of the critical
temperature for the exactly solvable case (for external magnetic field $H=0$) is similar to
the presented ones in this paper.
  
\section*{Acknowledgments}

This work has been supported by Slovak Grant Agency, Grant n. 2/4109/98.
We would like to thank the organizers of the DMRG Seminar and Workshop in Dresden 
for the opportunity to participate in the meetings, especially for the useful 
discussion with T.~Nishino.


\end{document}